\documentclass[pdflatex,sn-mathphys,iicol]{sn-jnl}

\jyear{2022}%

\theoremstyle{thmstyleone}%

\theoremstyle{thmstyletwo}%

\theoremstyle{thmstylethree}%

\usepackage{geometry,amsmath,amsfonts,amssymb,xcolor}
\begin{document}

\title[Article Title]{Spectral interactions between strings in the Higgs background}

\author*[1]{\fnm{Arkadiusz} \sur{Bochniak}}\email{arkadiusz.bochniak@doctoral.uj.edu.pl}

\author[1]{\fnm{Andrzej} \sur{Sitarz}}\email{andrzej.sitarz@uj.edu.pl}

\affil*[1]{\orgdiv{Institute of Theoretical Physics}, \orgname{Jagiellonian University}, \orgaddress{\street{prof. Stanis{\l}awa {\L}ojasiewicza 11}, \city{ Krak{\'o}w}, \postcode{30-348}, \country{Poland}}}

\abstract{We derive the exact form of the spectral interaction of two strings mediated by a constant scalar field using methods derived from noncommutative geometry. This is achieved by considering
a non-product modification of the Connes-Lott model with two-dimensional manifolds. The analogy with the latter construction justifies the interpretation of the scalar field as being of Higgs type.
Working in dimension two requires the use of the spectral zeta function instead of the Wodzicki residue techniques applicable to four-dimensional models. In the latter case, an analogous non-product
geometry construction leads, for specific choices of metrics, to the so-called "doubled geometry models", which can be thought of as a spectral modification of the Hassan-Rosen bimetric theory. We find that
in dimension two, the interaction term depends explicitly on zweibeins defining the Dirac operators and only in some special cases can they be expressed solely using the metrics. The computations can be
performed analytically for an arbitrary choice of zweibeins defining geometry on the two strings.}

\keywords{Noncommutative Geometry, Bimetric Theory, Spectral Action}



\maketitle

\section{Introduction}
	
	The spectral methods of noncommutative geometry \cite{Con95,Con96} can be successfully applied to describe both General Relativity \cite{Kas95,Con93} and Standard Model of particle physics together with its potential extensions \cite{ConnesLott, DunSu12, SuBook, Con90}. The fundamental object in noncommutative geometry, a spectral triple, is motivated by the observation made by A.~Connes that for any sufficiently regular Riemannian manifold $M$ that allows spinor fields, one can associate a triple consisting of an algebra $A=C^\infty(M)$ of smooth functions on $M$ 
   represented on a Hilbert space $H=L^2(S)$ of square-integrable 
   spinors and the Dirac operator given locally as $\mathcal{D}=i\gamma^\mu (\partial_\mu +\omega_\mu)$, where $\omega$ is the spin connection on $M$ and $\gamma$'s are the usual generators of the associated Clifford algebra. By Connes' reconstruction theorem \cite{Con96,Con13}, any suitably regular triple $(A,H,\mathcal{D})$, with an abelian algebra $A$, is of the above {\it canonical} form. This motivates the notion of a spectral triple as a basic object that generalizes the notion of geometry to objects that are discrete spaces, fractals, or that are described by algebras, which are no longer assumed to be commutative. The data of a spectral triple consists of a $\ast$-algebra $A$ represented (faithfully) on a Hilbert space $H$ on which $\mathcal{D}$ is as an essentially self-adjoint operator having compact resolvent, such that its commutators with elements of the algebra $A$ are bounded. A set of possible additional structures, like grading and real structure, can be incorporated into this picture and additional compatibility conditions between all these elements can be assumed. For a more detailed discussion see e.g. \cite{Con94, Li17}. 
	
Out of the above spectral data, one can compute the so-called spectral action \cite{EI18} which is essentially a functional in $\mathcal{D}$. By spectral action principle, its leading terms in heat kernel expansion, \cite{heat} describe the physical model obtained from the geometry associated with a given spectral triple. This generalizes Einstein's idea of equivalence between geometry and physics (gravity). Indeed, the computation of the (leading terms of) the spectral action for the canonical spectral triple associated with a given manifold leads to the Hilbert-Einstein action. 
    
This picture was expanded to include Yang-Mills gauge theories within the framework of almost-commutative geometry when the spectral triple is a product of the canonical one and another one with both an algebra and a Hilbert space being finite-dimensional  \cite{ Con95, Con96}. Appropriate choice of the finite part (closely related to the gauge group one is interested in) allowed for a formulation of the Standard Model of particle physics within this framework and to study its properties as well as symmetries \cite{lepto99, DaAnSi, AnKuLi}.  
	
	However, the almost-commutative framework seems not to be the final answer for all physically relevant questions. In particular, it is very restrictive in the gravitational sector and it leads (in its bare version) to some unphysical behavior in particle physics models. Several non-product modifications of almost-commutative geometries were therefore proposed. Some of them are formulated at the level of Kasparov modules \cite{BoevS,BoevD}, whereas others are direct modifications motivated by potential applications both for the Standard Model \cite{BoSi20, BoSiZa21} as well as for modified gravity theories. In \cite{Si19}, an approach based on the straightforward modification of the Connes-Lott model \cite{ConnesLott} was proposed. The basic idea was to replace the almost-commutative geometry $M\times \mathbb{Z}_2$ by a direct sum of two copies of $M$ but considered with different metrics on each summand. This so-called doubled geometry model was further studied in the case of Friedmann-Lema\^itre-Robertson-Walker metrics in \cite{BoSi21} and for other classes of metrics also in \cite{Bo22}. Since the resulting action contains interactions between the two metrics it is natural to ask about potential relations to Hassan-Rosen bimetric gravity theories \cite{HaRo12, Akrami13}. We have discussed partially this aspect in \cite{BoSi21} and continued in \cite{BoSi22}, where the picture of two four-dimensional branes interacting effectively by Higgs-like potential was introduced. 
	
Let us briefly summarize main steps in the construction of the doubled models with two copies of the spin manifold $M$ of dimension $d>0$. As the Dirac operator, we take
\begin{equation}
		\mathcal{D}=\begin{pmatrix}
			\mathcal{D}_g & \gamma \Phi\\
			\gamma\Phi^\ast & \mathcal{D}_h
		\end{pmatrix},
	\end{equation}
	where $\mathcal{D}_g$ (resp. $\mathcal{D}_h$) is the standard Dirac operator corresponding to the Riemannian manifold $(M,g)$ (resp. $(M,h)$), $\Phi$ is a (constant) field and $\gamma$ is taken so that $\gamma^2=\pm 1$ and it anticommutes with the Dirac operators on both {\it sheets}. The computation of all relevant terms of the spectral action is, however, dimension 	dependent. The leading term can be expressed using the Wodzicki residue of 
	$\lvert\mathcal{D}\rvert^{-d}$ and will give the sum of volumes of each copy of $M$ (with respect to the metric $g$ and $h$, respectively). However, the next term, which will contain the sum of integrated scalars of curvature for each copy, will also have an interaction term between the two metric manifolds. In the case of $d=2$ the explicit computation of these terms necessitates the use of the zeta function \cite{ConTre, ZaKha}. 
	
	The tools to compute the corresponding terms of spectral action are based on the calculus of pseudodifferential operators. We first need to find the symbols of the operator $\mathcal{D}^2$ by replacing each derivative $\partial_j$ with a formal variable $i\xi_j$. Let $\mathfrak{a}_k$ denotes the part of this symbol that is homogeneous in $\xi_\mu's$ with the homogeneity degree $k$. In the next step, we compute the symbols $\mathfrak{b}_\bullet$ for $\mathcal{D}^{-2}$ \cite{Gilkey}, and it turns out that only two of them ($\mathfrak{b}_0$ and $\mathfrak{b}_2$) are enough to obtain two leading terms of the spectral action. 
	
    Motivated by the fact that this approach can produce spectral interactions of branes, we focus in this paper on the case of two-dimensional branes, the strings. Our model does not assume any ambient space in which the strings propagate as the action is fully intrinsic and arises from the scalar field that links the two strings. Another 
    possible interpretation is that of a ``thick string'' where the 
    resulting action is the interaction of its boundaries (surfaces) in 
    the Higgs background.
    
	We concentrate on the situation where the action comes from the generalized Dirac operator that involves the Higgs field. Thus, we work with metrics given in terms of zweibeins. In section \ref{sec:zwei} we formulate the two-dimensional Riemannian doubled geometry model in terms of zweibeins and derive the resulting effective action using spectral methods. Then, in section \ref{sec:met} we briefly discuss a reformulation of this model in terms of the corresponding 
    metrics. Finally, in section \ref{sec:fin} we shortly discuss the 
    Lorentzian formulation, interpretation, and present an outlook. 
	
	\section{Formulation in terms of zweibeins}
	\label{sec:zwei}
	For a two-dimensional Riemannian spin${}_c$ manifold with a given metric $g$ defined in terms of a zweibein $e^\mu_a$ (with an inverse metric $g^{\mu\nu}=e^\mu_a e^\nu_b \delta^{ab}$) we consider the Dirac operator of the form
	\begin{equation}
		\mathcal{D}_g=i\sigma^a e^\mu_a \partial_\mu -\frac{1}{2}\sigma^a \partial_\mu (e^\mu_a),
	\end{equation}
	with both $a,b=1,2$ and $\mu=1,2$, and the Pauli matrices $\sigma^a$ satisfying $\{\sigma^a,\sigma^b\}=2\delta^{ab}I$. Let 
	$$K=\begin{pmatrix}a_1& d_1\\ c_1&b_1\end{pmatrix}, $$ 
	be the matrix corresponding to the zweibein $e^\mu_a$, i.e. with entries $K^{\mu}_{\ a}:=e^{\mu}_a$. Since there is no one-to-one correspondence between metrics and zweibeins (a zweibein determines the metric but a metric can	correspond to a class of zweibeins), in this paper, we treat the zweibein $e^\mu_a$, rather than the metric $g$, as a more fundamental object.
	
In the doubled model we denote the second metric by $h$, the 
second zweibein by $f^\mu_a$ and its corresponding matrix by $L$.
	
	Since we are interested mostly in the form of the interaction potential we can omit terms with derivatives of zweibeins. Under this assumption, the Dirac operator for the doubled geometry model takes the form
	\begin{equation}
		\mathcal{D}=i\sigma^a A^\mu_a\partial_\mu +\sigma F,
	\end{equation}
	with 
	\begin{equation}
		A^\mu_a=\begin{pmatrix}
			K^{\mu}_{\ a} &  0\\
			0& L^{\mu}_{\ a}
		\end{pmatrix}, \quad F= \begin{pmatrix}0 & \Phi\\ \Phi^\ast & 0\end{pmatrix},
	\end{equation}
	where $\Phi$ is assumed to be a constant, and $\sigma=\chi \sigma^3$ with $\chi$ chosen s.t. $\sigma^2=\kappa=\pm 1$.
	We can now easily compute the operator $\mathcal{D}^2$, and its symbols read
	\begin{equation}
		\begin{split}
			&\mathfrak{a}_2=\begin{pmatrix}
				(K\xi)^2 & 0\\
				0 & (L\xi)^2
			\end{pmatrix}=\begin{pmatrix}\|\xi\|_g^2 & 0 \\ 0& \|\xi\|_h^2\end{pmatrix},\\
			&\mathfrak{a}_1=[F,A^\mu_a]\sigma^a\sigma \xi_\mu,\\
			&\mathfrak{a}_0=\kappa F^2,
		\end{split}
	\end{equation}
	where $\xi=\begin{pmatrix}\xi_1\\ \xi_2\end{pmatrix}$ is the vector of symbols of derivatives (i.e. $\partial_j\leftrightarrow i\xi_j$ for $j=1,2$).

	The computation of the symbols of $\mathcal{D}^{-2}$ simplifies for the derivative-free case and we get \cite{Gilkey}:
	\begin{equation}
		\mathfrak{b}_0=(\mathfrak{a}_2+1)^{-1},\qquad
		\mathfrak{b}_2=-\mathfrak{b}_0\mathfrak{a}_0\mathfrak{b}_0+\mathfrak{b}_0\mathfrak{a}_1\mathfrak{b}_0\mathfrak{a}_1\mathfrak{b}_0. 
	\end{equation}
	We remark that the above form of the $\mathfrak{b}_0$ symbol is a consequence of the fact that the manifold is two-dimensional \cite{ConTre, ZaKha}.
	
	To proceed with the computation of the spectral action, we first notice that  
	\begin{equation}
 \begin{split}
		&\mathrm{Tr\ }\mathrm{Tr}_{Cl}(\mathfrak{b}_0^2)\\
  &=2\left(\frac{1}{((K\xi)^2+1)^2}+\frac{1}{((L\xi)^2+1)^2} \right)
  \end{split}
	\end{equation}
	and 
	\begin{equation}
		\int \frac{d^2\xi}{((K\xi)^2+1)^2}=\frac{\pi}{\det(K)},
	\end{equation}
	where the trace $\mathrm{Tr}$ is over the discrete degrees of freedom and $\mathrm{Tr}_{Cl}$ is the trace over the Clifford algebra. Then, the contribution to the action from this term is simply 
	$$ 2\pi\left(\frac{1}{\det(K)}+\frac{1}{\det(L)}\right),$$ 
	and gives the standard cosmological constant terms for the two metrics.
	
	For the next term in the action, we obtain first,
	\begin{equation}
		\begin{split}
			&\mathrm{Tr\ }\mathrm{Tr}_{Cl}(-\mathfrak{b}_0\mathfrak{a}_0\mathfrak{b}_0)\\
      &=-2\kappa\lvert\Phi\rvert^2\left[\frac{1}{\left((K\xi)^2 +1\right)^2}+\frac{1}{\left((L\xi)^2 +1\right)^2}\right]\\
			&=-2\pi \kappa \lvert\Phi\rvert^2 \left(\frac{1}{\det(K)}+\frac{1}{\det(L)}\right),
		\end{split}    
	\label{eq9}
	\end{equation}
	which is just a correction to the previous part. 
	
	Finally, the integrand of the remaining part of the potential is, 
	\begin{equation}
 \begin{split}
		&\mathrm{Tr\ }
		\mathrm{Tr}_{Cl}(\mathfrak{b}_0\mathfrak{a}_1\mathfrak{b}_0\mathfrak{a}_1\mathfrak{b}_0)\\
  &=-2\kappa\mathfrak{b}_0[F,A^\mu_a]\mathfrak{b}_0[F,A^\nu_b]\mathfrak{b}_0\delta^{ab}\xi_\mu\xi_\nu,
  \end{split}
	\end{equation}
	which gives 
	\begin{equation}
 \begin{split}
		&\mathrm{Tr\ }
		\mathrm{Tr}_{Cl}(\mathfrak{b}_0\mathfrak{a}_1\mathfrak{b}_0\mathfrak{a}_1\mathfrak{b}_0)\\
  &=2\kappa \lvert\Phi\rvert^2\det(\mathfrak{b}_0)\mathrm{Tr}(\mathfrak{b}_0)((K-L)\xi)^2,
  \end{split}
	\end{equation}
	and, as a result, we obtain the following integration formula
	\begin{equation}
 \begin{split}
 \label{eq:contr}
		2\kappa\lvert\Phi\rvert^2\int d^2\xi &\left( \frac{((K-L) \xi)^2}{((K\xi)^2+1)^2 ((L\xi)^2+1)} \right. \\
  &\left.
		+ \frac{((K-L) \xi)^2}{((K\xi)^2+1) ((L\xi)^2+1)^2} \right).
  \end{split}
	\end{equation}
	
	Let us denote the above contribution \eqref{eq:contr} to the potential by $V_1$. In order to compute this integral, notice that the second term is just obtained from the first one by exchanging $K$ with $L$. So, for a moment we concentrate on the first term only. Let then $S$ be an orthogonal matrix that diagonalizes 
	$(K^{-1})^T L^T L K^{-1}$ to $\Lambda^T \Lambda$, i.e. 
	\begin{equation}
		\label{eq:lambda}
		S^{-1}(K^{-1})^T L^T L K^{-1} S=\Lambda^T\Lambda,
	\end{equation}
	and take $P=K^{-1} S$. Here we have used the fact that $(K^{-1})^T L^T L K^{-1}$ have positive eigenvalues. 
	
	Then changing the coordinates according to $P \eta = \xi$ we obtain
	\begin{equation}
 \begin{split}
 & \det(K^{-1} S) \int d^2\eta 
		\frac{ ((S - L K^{-1} S) \eta)^2}{(\eta^2+1)^2 ((\Lambda \eta)^2+1)}\\
  & \equiv \det(K^{-1} S) \int d^2\eta 
		\frac{ (W \eta)^2}{(\eta^2+1)^2 ((\Lambda \eta)^2+1)},
  \end{split}
	\end{equation}
	where we have introduced $W=S - L K^{-1} S$.
	
	As the denominator is symmetric with respect to $\eta$,
	only the parts with $\eta_1^2, \eta_2^2$ in the numerator
	will enter, so the considered part of the potential reads
	\begin{equation}
		\begin{split}
			&V_1=2\kappa\lvert\Phi\rvert^2 \det(K^{-1}S)\\
   &\times\int d^2\eta \left((W_{11}^2+W_{21}^2)\eta_1^2+(W_{22}^2+W_{12}^2)\eta_2^2\right)\\
			\times&\left(\frac{1}{(\eta^2 +1)^2\left((\Lambda_1 \eta_1)^2+(\Lambda_2\eta_2)^2+1\right)}\right.\\
   &\left.+\frac{1}{(\eta^2 +1)\left((\Lambda_1 \eta_1)^2+(\Lambda_2\eta_2)^2+1\right)^2}\right),
		\end{split}
	\end{equation}
	where we used the fact that $(W\eta)^2=\eta^T W^TW\eta$ is symmetric with respect to the interchange $K\leftrightarrow L$. This can be written as
	\begin{equation}
		\begin{split}
			2\kappa\lvert\Phi\rvert^2 \det(K^{-1}S)&\left[(W_{11}^2+W_{21}^2) I_1(1,1,\Lambda_1,\Lambda_2)\right. \\
   +&\left. (W_{22}^2+W_{12}^2) I_1(1,1,\Lambda_2,\Lambda_1)\right.\\ +&\left. (W_{11}^2+W_{21}^2) I_1(\Lambda_1,\Lambda_2,1,1)\right. \\ +&\left.(W_{22}^2+W_{12}^2) I_1(\Lambda_2,\Lambda_1,1,1)\right],
		\end{split}    
	\end{equation}
	where
	\begin{equation}
		\begin{split}
			&I_1(a,b,c,d)\\
   &= \int_{\mathbb{R}^2}\frac{\xi_1^2 d\xi_1d\xi_2}{(a^2\xi_1^2 + b^2\xi_2^2+1)^2(c^2\xi_1^2+ d^2\xi_2^2+1)}\\
   &=\frac{c}{a}\frac{\pi}{(c^2-a^2)(bc+ad)}+	\frac{\pi}{(c^2-a^2)^{3/2}(b^2-d^2)^{1/2}}\\
   &\hspace{2.5cm}\times\left[\arcsin\left(\frac{a\sqrt{b^2-d^2}}{\sqrt{b^2c^2-a^2d^2}}\right)\right.\\
   &\hspace{2.5cm}\left.-\arcsin\left(\frac{c\sqrt{b^2-d^2}}{\sqrt{b^2c^2-a^2d^2}}\right)\right].
		\end{split}
	\end{equation}
	Since
	\begin{equation}
 \begin{split}
		I_1(1,1,\Lambda_1,\Lambda_2)+I_1&(\Lambda_1,\Lambda_2,1,1)\\
  &=\frac{\pi}{\Lambda_1+\Lambda_2} \,\frac{1}{\Lambda_1},
  \end{split}
	\end{equation}
	we finally get
	\begin{equation}
		\begin{split}
			V_1&=\frac{2\pi \kappa\lvert\Phi\rvert^2 \det(K^{-1}S)}{\Lambda_1+\Lambda_2 }\\
   &\times\left(\frac{W_{11}^2+W_{21}^2}{\Lambda_1}+\frac{W_{22}^2+W_{12}^2}{\Lambda_2}\right) \\
			&=\frac{2\pi \kappa\lvert\Phi\rvert^2 \det(K^{-1}S)}{\mathrm{Tr}(\Lambda)}\mathrm{Tr}(W^TW\Lambda^{-1}).
		\end{split} 
	\end{equation}
	But since $S$ is an orthogonal matrix, this expression can be further reduced to
	\begin{equation}
		\label{eq:final}
		2\pi \kappa\lvert\Phi\rvert^2 \frac{\mathrm{Tr}(W^TW \Lambda^{-1})}{\mathrm{Tr}(\Lambda)\det(K)}.
	\end{equation}
	Let us now express \eqref{eq:final} in terms of invariants of the matrix $X=LK^{-1}$. First, by \eqref{eq:lambda} we have
	\begin{equation}
		\mathrm{Tr}(\Lambda)=\mathrm{Tr}((X^T X)^{1/2}).
	\end{equation}
	
	Next, again by \eqref{eq:lambda} and the fact that $X^TX$ is positive (hence $\Lambda^T\Lambda$ has positive eigenvalues), we have
	\begin{equation}
 \begin{split}
		\Lambda^{-1}&=(S^{-1}X^T X S)^{-1/2}=S^{-1}(X^TX)^{-1/2}S\\
  &=S^{T}(X^TX)^{-1/2}S,
  \end{split}
	\end{equation}
	and therefore
	\begin{equation}
		\begin{split}
			W^TW\Lambda^{-1}&=S^T(X^TX)^{-1/2}S\\ &-S^TX(X^TX)^{-1/2}S \\&-S^TX^T(X^TX)^{-1/2}S\\
   &+S^T(X^TX)^{1/2}S,
		\end{split}
	\end{equation}
	so that
	\begin{equation}
		\begin{split}	
			\mathrm{Tr}(W^TW \Lambda^{-1})&=
			\mathrm{Tr}((X^TX)^{-1/2})\\
   &-\mathrm{Tr}(X(X^TX)^{-1/2})
			\\ &-\mathrm{Tr}(X^T(X^TX)^{-1/2})\\ &+\mathrm{Tr}((X^TX)^{1/2}).
		\end{split}    
	\end{equation}
	Moreover, since $X^TX$ has only positive eigenvalues, we also have
	\begin{equation}
		\begin{split}
			(X(X^TX)^{-1/2})^T&=((X^TX)^{-1/2})^T X^T\\ &=\left[\left((X^TX)^{1/2}\right)^T\right]^{-1}X^T \\
			&=\left[\left((X^TX)^T\right)^{1/2}\right]^{-1}X^T\\ &=(X^TX)^{-1/2}X^T
		\end{split}    
	\end{equation}
	and therefore,
	\begin{equation}
 \begin{split}
		\mathrm{Tr}\left(X^T(X^TX)^{-1/2}\right)&=\mathrm{Tr}\left((X(X^TX)^{-1/2})^T\right)\\
  &=\mathrm{Tr}(X(X^TX)^{-1/2}),
	\end{split}
 \end{equation}
	so that
	\begin{equation}
 \begin{split}
		\mathrm{Tr}(W^TW \Lambda^{-1})&=\mathrm{Tr}((X^TX)^{-1/2})\\ &-2\,\mathrm{Tr}(X(X^TX)^{-1/2})\\ &+\mathrm{Tr}((X^TX)^{1/2}).
  \end{split}
	\end{equation}
	As a result,
	\begin{equation}
 \begin{split}
		&V_1=\frac{2\pi \kappa\lvert\Phi\rvert^2}{\det(K)} \cdot \frac{1}{\mathrm{Tr}((X^T X)^{1/2})}\\
  &\times \left[\mathrm{Tr}((X^TX)^{-1/2})
  -2\,\mathrm{Tr}(X(X^TX)^{-1/2}) \right. \\
  &\left.\hspace{0.5cm} +\mathrm{Tr}((X^TX)^{1/2})\right].
  \end{split}
  \end{equation}
	Since for the $2\times 2$ matrix $(X^TX)^{1/2}$ we have  
	\begin{equation}
  \label{eq:2by2trace}
	\begin{split}
 \mathrm{Tr}((X^TX)^{-1/2})&=\frac{\mathrm{Tr}((X^TX)^{1/2})}{\det((X^TX)^{1/2})}\\
 &=\frac{\mathrm{Tr}((X^TX)^{1/2})}{\lvert\det(X)\rvert},
	\end{split}
 \end{equation}
	this expression can be further rewritten as
	\begin{equation}
 \begin{split}
		&V_1=\frac{2\pi \kappa\vert\Phi\rvert^2}{\det(K)} \\ &\times\left(1+\frac{1}{\lvert\det(X)\rvert}-2\frac{\mathrm{Tr}(X(X^TX)^{-1/2})}{\mathrm{Tr}((X^T X)^{1/2})}\right).
  \end{split}
	\end{equation}
	
	In order to proceed further, we use the fact \cite{Levinger} that for a $2\times 2$ positive matrix $Y$ we have 
	\begin{equation}
		\sqrt{Y}=\frac{Y +\sqrt{\det(Y)}I}{\sqrt{\mathrm{Tr}(Y)+2\sqrt{\det(Y)}}},
	\end{equation}
	and hence
	\begin{equation}
		\mathrm{Tr}(\sqrt{Y})=\sqrt{\mathrm{Tr}(Y)+2\sqrt{\det(Y)}}.
	\end{equation}
	Furthermore,
	\begin{equation}
		\begin{split}
			&Y^{-1/2}=\sqrt{Y^{-1}}\\
   &=\frac{Y^{-1} +\sqrt{\det(Y^{-1})}I}{\sqrt{\mathrm{Tr}(Y^{-1})+2\sqrt{\det(Y^{-1})}}}\\
			&=\frac{Y^{-1}+\frac{1}{\sqrt{\det(Y)}}I}{\sqrt{\frac{\mathrm{Tr}(Y)}{\det(Y)}+\frac{2}{\sqrt{\det(Y)}}}}\\
			&=\frac{\sqrt{\det(Y)}Y^{-1}+I}{\sqrt{\mathrm{Tr}(Y)+2\sqrt{\det(Y)}}}.
		\end{split}    
	\end{equation}
	Therefore,
	\begin{equation}
  \mathrm{Tr}(X Y^{-1/2})=\frac{\sqrt{\det(Y)}\mathrm{Tr}(XY^{-1})+\mathrm{Tr}(X)}{\sqrt{\mathrm{Tr}(Y)+2\sqrt{\det(Y)}}}
	\end{equation}
	and
	\begin{equation}
		\frac{\mathrm{Tr}(X Y^{-1/2})}{\mathrm{Tr}(Y^{1/2})}=\frac{\sqrt{\det(Y)}\mathrm{Tr}(XY^{-1})+\mathrm{Tr}(X)}{\mathrm{Tr}(Y)+2\sqrt{\det(Y)}}.
	\end{equation}
	But in our case $Y=X^TX$, so that $XY^{-1}=X(X^TX)^{-1}=XX^{-1}(X^T)^{-1}=(X^T)^{-1}$ and hence $\mathrm{Tr}(XY^{-1})=\mathrm{Tr}(X^{-1})=
	\frac{\mathrm{Tr}(X)}{\det(X)}$. Furthermore, $\sqrt{\det(Y)}=\lvert \det(X)\rvert$ and therefore
\begin{equation}
		\frac{\mathrm{Tr}(X Y^{-1/2})}{\mathrm{Tr}(Y^{1/2})}=\frac{(1+\mathrm{sgn}(\det(X)))\,\mathrm{Tr}(X)}{\mathrm{Tr}(X^TX)+2\lvert\det(X)\rvert}.
	\end{equation}

For $\det(X)<0$ this term vanishes identically. On the other hand, for $\det(X)>0$ we get 

	\begin{equation}
 \begin{split}
		V_1&=\frac{2\pi \kappa\lvert \Phi\rvert^2}{\det(K)}\\
  \times&\left(1+\frac{1}{\det(X)}-\frac{4\,\mathrm{Tr}(X)}{\mathrm{Tr}(X^TX)+2 \det(X)}\right).
  \end{split}
  \label{eq:V1F}
	\end{equation}

Observe that the first two terms are, in fact, again of the same type as \eqref{eq9}, however, with an opposite sign. Therefore, in the end, the additional term proportional to the volumes of the manifolds that is proportional to $\lvert\Phi\rvert^2$ vanishes due to this cancellation. What remains as an effective interaction between the zweibeins on the two worldsheets is
	\begin{equation}
	V_{\mathrm{int}}= - \frac{8\pi \kappa\lvert\Phi\rvert^2}{\det(K)} \left(\frac{\mathrm{Tr}(X)}{\mathrm{Tr}(X^TX)+2\det(X)}\right).
\end{equation}
This can be further simplified using the fact that
$$ \det(X) = \frac{1}{2} \left( \mathrm{Tr}(X)^2 - \mathrm{Tr}(X^2) \right).$$

Note that although the expression above uses $K$ and $X$ it can be equivalently rewritten using $L$ and $X^{-1}$ as it is completely symmetric.

We remark that the above cancellation was possible due to the fact that we are working in two dimensions. Indeed, the first term in \eqref{eq:V1F} has its form because for a $2\times 2$ matrix the trace of its inverse can be expressed in terms of the trace of the original matrix and its determinant, as it was done in Eq.\eqref{eq:2by2trace}. As a result, in two dimensions, the contribution to the action from $\mathfrak{b}_2$ does not affect the effective cosmological constant, in contrast to dimension four \cite{BoSi21}. Moreover, the interaction is non-zero only if $\det(X)$ is positive, i.e., both $K$ and $L$ have determinants of the same signs.

Before we discuss the dependence of the above term on the metrics let us consider
a special case, of diagonal zweibeins.

\subsection{A special case: diagonal constant zweibeins}
	
We consider now the diagonal case, i.e. with $K=\mathrm{diag}(a_1,b_1)$ and $L=\mathrm{diag}(a_2,b_2)$.  In such a situation we have $$
X=\mathrm{diag}\left(\frac{a_2}{a_1},\frac{b_2}{b_1}\right),$$
so that
	\begin{equation}
 \begin{split}
		\mathrm{Tr}(X)=\frac{a_1b_2+a_2b_1}{a_1b_1},\quad &\mathrm{Tr}(X^TX)=\frac{a_1^2b_2^2+a_2^2b_1^2}{a_1^2b_1^2},\\ \det(X)=&\frac{a_2b_2}{a_1b_1},
  \end{split}
	\end{equation}
	and therefore if we assume that $\det(X)>0$ then
	\begin{equation}
		V_{\mathrm{int}}=-\frac{8\pi\kappa \lvert\Phi\rvert^2}{a_1b_2+a_2b_1}.
	\end{equation}

\section{Formulation in terms of metrics}
\label{sec:met}
The derivation of the interaction terms between the two worldsheets was expressed
in the previous section directly in terms of zweibeins, which define the Dirac operators.
However, it is interesting to see which part of it can be rewritten explicitly in terms
of the metrics.  In this section we discuss this possibility, using the metrics $g,h$ on 
the two layers of the doubled model. 

First of all, observe that if a zweibein is given by the matrix
	\begin{equation}
	K=\begin{pmatrix}
		a_1 & d_1 \\
		c_1 &  b_1
	\end{pmatrix},
\end{equation}	
then we have
	\begin{equation}
		g^{-1}=\begin{pmatrix}
			a_1^2+d_1^2 & a_1c_1+b_1d_1 \\
			a_1c_1+b_1d_1 & b_1^2+c_1^2
		\end{pmatrix},
	\end{equation}
and the metric
	\begin{equation}
 \begin{split}
		g=&\frac{1}{(a_1b_1-c_1d_1)^2}\\
  &\times \begin{pmatrix}
			b_1^2+c_1^2 & -(a_1c_1+b_1d_1)\\
			-(a_1c_1+b_1d_1) & a_1^2+d_1^2
		\end{pmatrix}.
  \end{split}
	\end{equation}
	Therefore,
	\begin{equation}
		\det(K)=a_1b_1-c_1d_1=\sqrt{\det(g^{-1})},
	\end{equation}
	and similarly for the matrix $L$ and the metric $h$.
	Next,
	\begin{equation}
		X=\sqrt{\det(g)}\begin{pmatrix}
			a_2 b_1 -c_1d_2 & a_1d_2 -a_2d_1\\
			b_1c_2-c_1b_2 & a_1b_2-d_1c_2
		\end{pmatrix}
	\end{equation}
	and 
	\begin{equation}
 \begin{split}
		\det(X)&=\frac{\det(L)}{\det(K)}=\frac{\sqrt{\det(h^{-1})}}{\sqrt{\det(g^{-1})}}\\ &=\sqrt{\det(gh^{-1})}
  =\det\left(\sqrt{gh^{-1}}\right).
	\end{split}
 \end{equation}
 	
Therefore, the only obstruction to rephrasing the action in terms of metrics rather than zweibeins is given by the term that contains a trace of the matrix $X$ as well as the trace of $X^TX$.

If the zweibeins are such that these quantities are expressible in terms of invariants of $g$, $h$, and/or $gh^{-1}$, then the full action can be unambiguously written on the level of metrics. However, in the most general case, we are dealing with interactions between zweibeins rather than metrics. This is in contrast to the standard bimetric theory, which was formulated from the beginning using metrics. The vielbein version was shown \cite{HiRo} to be dynamically equivalent to the usual bimetric theory, however, that is not necessarily true for multi-metric models (see \cite{HaMay}). Of course, this is a consequence of the assumed model of interactions between the two vielbeins and the basic structure where spinor fields and vielbeins are fundamental objects. It is interesting whether there exists a model of the full Dirac operator, where the effective potential depends only on the metrics.

\subsection{A special case: diagonal constant zweibeins}
	
In the case of diagonal zweibeins, from the form of the principal symbol, we easily read that the metrics are such that their inverses are of the form $\mathrm{diag}(a^2,b^2)$, i.e., they are also diagonal. In this case $\sqrt{\det(g)}=\frac{1}{a_1 b_1}$ and $\sqrt{\det(h)}=\frac{1}{a_2b_2}$.

Let us consider the matrix $\mathbb{X}=\sqrt{g^{-1}h}$. Its eigenvalues are $x=\frac{a_1}{a_2}$ and $y=\frac{b_1}{b_2}$, and therefore,
	\begin{equation}
 \begin{split}
		V_{\mathrm{int}}&=-8\pi \kappa\lvert\Phi\rvert^2\sqrt{\det(g)}\underbrace{\frac{xy}{x+y}}_{V\left(\sqrt{g^{-1}h}\right)}\\
  &=-8\pi \kappa\lvert\Phi\rvert^2\sqrt{\det(h)}V\left(\sqrt{h^{-1}g}\right).
  \end{split}
	\end{equation} 
 
\section{Lorentzian formulation and conclusions}
\label{sec:fin}
Since the derivation and applicability of the spectral methods rely on a purely Riemannian framework, the final action describes a Riemannian model. However, we can use the standard Wick rotation argument to make contact with the Lorentzian formulation. The procedure is as follows. Since we started with the modified Connes-Lott model, the effective interaction originating from the field $\Phi$ has the character of a Higgs-like one. The two-dimensional Riemannian manifold we started with is now replaced by 
a $(1+1)$-dimensional string, and we end up with a model of two 
Lorentzian strings interacting via a Higgs-like field. The interaction is given at the level of corresponding zweibeins for these two strings. This is the two-dimensional generalization of the picture with four-dimensional 
branes we discussed previously in \cite{BoSi22} to make a comparison between doubled geometry models \cite{Si19, BoSi21} and Hassan-Rosen bimetric gravity theories \cite{HaRo12}. 
		
Contrary to the four-dimensional situation, in dimension two we compute the
zeta function rather than a residue, and the resulting potential can be computed analytically for generic zweibeins. The resulting interaction is non-zero only if the matrices corresponding to the zweibeins have determinants of the same signs, and is then expressed in terms of the invariants of the matrix $X=LK^{-1}$ with parts of the result expressible only in terms of metrics. This is analogous to the formulation of bimetric gravity in terms of vielbeins and metrics \cite{HiRo} and similar to certain choices for vielbeins that ensure the existence of real square root $\sqrt{g^{-1}h}$ \cite{DeMoZa}. We postpone the detailed discussion of this aspect for future research.  
		
\backmatter

\bmhead{Acknowledgments}
A.~B. thanks S.F.~Hassan for insightful discussions and for stating the question that motivated this study. The authors acknowledge the support of the National Science Centre, Poland, Grant No. 2020/37/B/ST1/01540.

\bmhead{Competing interests} The authors declare no competing interests.

\bmhead{Author contributions} Both authors contribute equally to the work as well as to the writing of the manuscript. 

\bmhead{Data availability} All data generated or analyzed during this study are included in the manuscript.

\end{document}